___

# Clustering Optimisation Techniques in Mobile Networks

Eleni Rozaki
School of Computer Science and Informatics
Cardiff University
Cardiff, United Kingdom
rozakie@cs.cardiff.ac.uk

*Abstract*—The use of mobile phones has exploded over the past years, abundantly through the introduction of smartphones and the rapidly expanding use of mobile data. This has resulted in a spiraling problem of ensuring quality of service for users of mobile networks. Hence, mobile carriers and service providers need to determine how to prioritise expansion decisions and optimise network faults to ensure customer satisfaction and optimal network performance. To assist in that decision-making process, this research employs data mining classification of different Key Performance Indicator datasets to develop a monitoring scheme for mobile networks as a means of identifying the causes of network malfunctions. Then, the data are clustered to observe the characteristics of the technical areas with the use of k-means clustering. The data output is further trained with decision tree classification algorithms. The end result was that this method of network optimisation allowed for significantly improved fault detection performance.

___________________________________________________*****___________________________________________________

## I. INTRODUCTION

Clustering is a process in which similar data are collected and group together into clusters. The goal of clustering is to reveal structures that are otherwise hidden within the data. Clustering is an unsupervised process, meaning that the process is performed with the use of algorithms. The algorithms that are created are designed to place data together based on metrics that are chosen by the users. Generally, the metrics are based on conditions or aspects that are of importance to the tasks performed by the users [1].

An important benefit of clustering is that it allows for the reuse of resources, which results in an increase in system capacity. Rather than needing separate system resources and capacity for each cluster, multiple clusters can operate at the same time and actually share resources with each other. Within MANETs (mobile ad hoc networks), two types of clustering can be used: simple clustering and enhanced clustering [2].

Within the category of simple clustering, there are four types of clustering:

**Lowest ID Clustering**: Every node in the network is given a unique ID, and each node communicates with its neighbors. Nodes that hear from nodes having a higher ID value are grouped into cluster heads [3].

**Connectivity Clustering**: All nodes broadcast the nodes from which they can hear. A node that is connected to most of the uncovered neighbor nodes, meaning nodes that do not belong to any clusters, are the cluster heads. If multiple nodes have the same level of connectivity, then the node with the lowest ID is selected as the cluster head [3].

**Cell Clustering**: This type of clustering is used in mobile IP. A region is divided into cells, and a group of those cells are clustered. If two cells have a high degree of hand off between them, then those cells may be clustered [3].

**Weighted Clustering**: Each node calculates its weight based on its characteristics. The node that has the highest weight becomes the cluster head [3].

With the category of enhanced clustering, there are four types of clustering that generally used:

**K-Cluster Approach**: A cluster is considered to be a subset of notes in a graph-based approach. There is a path from never node to every other node, and every node contains structure tables with the information of neighboring nodes [3].

**Hierarchical Clustering**: All clusters are connected, and have minimum and maximum size limits. In addition, two clusters should have low overlap and they should be stable. The clustering algorithm consists of the tree discovery and the cluster formation [3].

**Dominating Sets Clustering**: Clustering occurs using dominating nodes. All nodes have an initial color of white, and are changed to black when it becomes the dominating node. Its neighbors are changed to gray at this point and become part of the dominating cluster [3].

___

**Max-Min D Clustering**: This approach has data structures and functions that are based on a winning node, a sender node, a floodmax function in which each node broadcasts its winning value to neighboring nodes, a floodmin function in which each node selects its smallest value as the winning value, overtake function in which winning values are propagated to neighboring nodes, and node pairs in which a node occurs at least once as a winning node in terms of both floodmax and floodmin [3].

Optimising the traffic on a cellular network is of great importance to network operators because of the need to achieve the highest network performance possible given the large numbers of subscribers using such networks to transfer large amounts of data on a daily basis [4]. The Universal Mobile Telecommunications System (UMTS) is a network standard in which network performance evaluation is based on analysis of the key performance indicators (KPIs) using data mining techniques for the purpose of network planning and resource optimisation [5].

Every network provider uses a specific method or model by which to identify problems that can result in inadequate network optimisation [6]. While many models and methods of automated network optimisation have been proposed and studied within the academic literature, the use of the Waikato Environment for Knowledge Analysis (Weka) machine learning for the purpose of analysing the values of KPI alarms, setting up rules and decision tree procedures to identify and diagnose rules and the condition and limits of network optimisation faults as a means of automating defect management needs to received further attention and interest in future years. Weka is a machine learning language that can be used to apply learning algorithms to data in order to examine output and performance [7].

With the use of J48 Classifier in this investigation, the problem that is to be solved is the categorisation of the causes that affect symptoms (KPIs) when the user quality of service (QoS) is not achieved [8][9]. Automated optimisation processes require a heavy usage of the infrastructure of a network, which means that the process actually needs to provide efficient and reliable information by which to identify and classify the causes of a lack of network performance improvement. The ultimate goal is to achieve RF optimisation in the most efficient manner possible in order to ensure that network QoS is achieved from the perspective of the user [10].

Within this paper, a methodology for RF failure detection in UMTS networks is proposed. The proposed methodology is based on our previous research which focuses on a troubleshooting scheme that classifies the optimisation faults in critical, warning and norm status. [11]. We are looking for the classification of causes and symptoms (KPIs) which are obtained by measuring certain performance parameters at different network entities. In this investigation, GSM network RF performance evaluation is presented based on the possible group of causes of network malfunctions that represent the diagnosis of the symptoms (major KPIs). Therefore, the first stage of this experiment is to apply the decision tree classifier in order to identify the fault causes (diagnosis). Next, the proposed clustering methodology is applied in order to analyze the characteristics of the different technical areas of concern.

The goal of this work is twofold: to investigate optimisation rules and examine the analysis methods of characteristics of a set of KPIs related to mobile network data. The importance of the investigation described in this paper is to provide information and insights to mobile network operators to engage in better centralized planning and equipment management of their networks. In addition, this investigation was designed to help network operators to optimise load balancing on mobile networks through the accurate and efficient scheduling of network resources [12].

## II. CLUSTERING SYSTEM DESIGN

One of the problems with much of the past research on clustering is that the focus has been on single objective. Single algorithms have been used to attempt to achieve network optimisation. Single objective algorithms cannot address multiple objectives, which is why multi-objective optimisation is required. The key to achieving multi-objective optimisation is to minimize the backbone by minimizing cluster heads [2].

In addition to minimizing the cluster heads, optimisation requires balancing the number of nodes that are covered by each of the cluster heads. Otherwise, a cluster head with too many nodes may collide with each other as they attempt to send and receive messages. At the same time, the power consumption of each node must be minimized. The power consumption of a node is based on the distance between two nodes that communicate with each other. In this regard, the distance between nodes that exchange information must be reduced so that power consumption is reduced, and minimized [2].

The data mining process is the basis for the detection of causes and problems for automated network optimisation. The data mining process consists of cleaning the data to remove irrelevant data, the integration of multiple data sources into a single dataset, the selection of the data that will actually be used in the analysis, and the transformation of the data into forms that are appropriate for actual data mining. Consequently, the data mining can begin, which involves the evaluation of patterns in the data. Finally, knowledge representation occurs in which visualisation techniques are used to place the knowledge gained from the data mining into visual elements that help users to both understand and interpret results and findings [13].

With the use of the Weka tools, the data mining process begins by defining the class attributes that divide the set of instances into appropriate classes or causes. Then, the selected sets of KPIs are defined and the causes are categorized in groups. Therefore, the relations and interdependencies of the KPIs are selected, followed by an investigation of any possible imbalance in the selected data set in order to determine how the imbalance may be counteracted [14].

The fault detection applications in Weka were discussed in the first stage of the plan to demonstrate the faults, symptoms, and causes of network issues. The tree classification in Weka perform the data for the optimisation algorithm. The primary center of classes in the iterative optimisation algorithm are chosen based on random selection. It should be noted that the use of random selection does result in the iterative time to be greatly increased. Furthermore, there are problems with this approach in terms of blindly selecting samples, omitting clustering tendency of the samples, and presenting local extremum [14].

Once the steps of defining the class attributes, extraction of the features to be used, and selection of the relations and interdependencies has occurred along with the investigation of any imbalances in the data, then the application of a classifier algorithm for the learning process can occur.

Then, a decision is made about the testing method that will be used to estimate performance related to a selected tree algorithm. Finally, the optimal solution to provide QoS for users is defined and implemented [9] [17].

### III. DATA SET DESCRIPTION

#### A. Defining the Relations Between the Symptoms (KPIs) and Causes.

The work of data preparation begins with analysing and pre-processing of database features, as well as an assessment of the correctness of the data that are to be analysed. The process of data preparation involves determining the specific data that can be used to identify network malfunctions. The selection of the data to be analysed involves several considerations, such as which data are representative of the KPIs (symptoms), data availability (inputs), the attributes of the data, and establishing limits of the network faults [17][18].

The data preparation performs the analysis and pre-processing of input data that are the symptoms and must be defined, such as dropped call rate (DCR), call success rate (CSR), traffic rate (TR), and stand-alone dedicated control channel success rate (SDCCHSR). The KPIs are extracted from data related to call attempts (Seizure) (CA), call failures (Seizure) (CF), call success (Seizure) (CS), incoming traffic (TE), outgoing traffic (OE), SDCCH seizure attempts (SDCCHSA), and successful Standalone Dedicated Control Channel seizures (SSDCCH) [15] [16].

In addition, the algorithm parameters must be set for the input data. The parameters will include equality or inequality for nominal relations and greater than or less than for number relations. The relational nodes can be placed into decisions trees or into rules. However, plans that use relations are more often placed in rules rather than in decision trees [6].

Fig. 1 shows the input of the optimisation system with the direct interdependencies of the KPIs and the causes. The abbreviations of the KPIs are as follows: CSSR is Call Setup Success Rate, TR is Traffic Rate, DCR is Dropped Call Rate, and SDCCHSR is Stand-Alone Dedicated Control Channel Success Rate. With the use of Fig. 2 it is possible to see the interdependencies of the KPIs and the groups of causes, such as GCA which refers to optimisation errors related to Call Setup Success and Dropped Call Rate. The GCB refers to traffic issues, while GCC needs to be classified based on the symptoms [7][11][21].

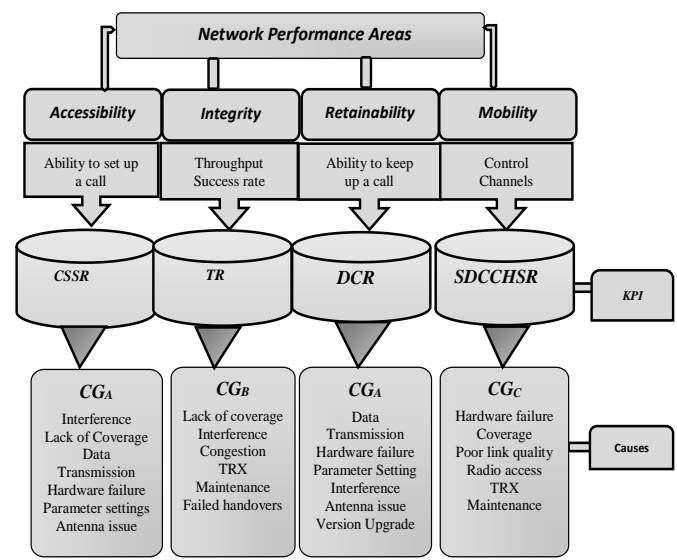

Figure 1: Direct independencies of KPIs and the causes

In order to run the data that were collected and to test the data performance, the relations of the attributes were setup using the C4.5 algorithm. Moreover, network fault limits were established for KPI relational nodes [16]. Based on those relational nodes, the system would identify the optimised status of the TCH (traffic channel), Handovers, RAB (Radio Access Barrier) and Traffic Channel Congestion Rate KPI's, and show them in their optimised states [22].

## B. Data Classification via Clustering

Clustering is used in many fields as a means of classifying knowledge. In this investigation, the detection of network faults occurs through the use of data mining techniques. In the mobile network, two important parts of the process of detecting network faults are performance management and faults management. Data can be obtained through several methods, including SNMP protocol, hardware analytical instruments such as HP WAN/LAN, and software analytical instruments. For this investigation, the data were obtained through SNMP protocol.

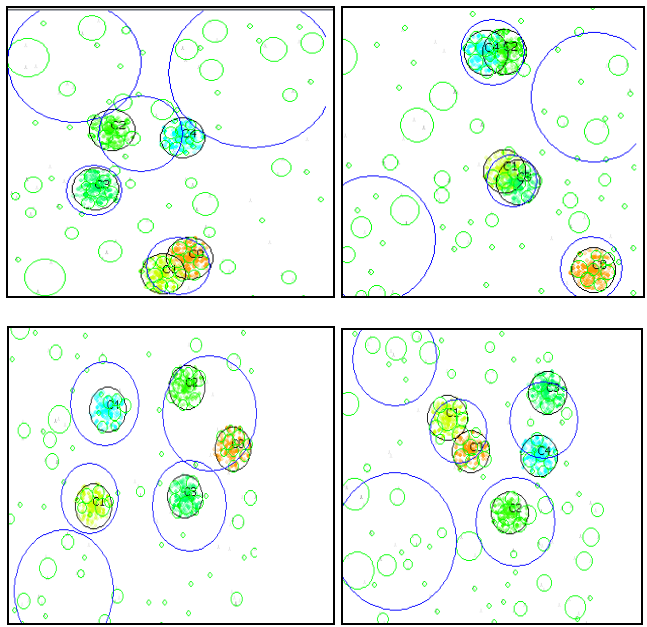

Figure 2: Clustered Optimisation Causes

The figure 2 shows the clusters range illustrate the different areas that categorise the k-means of the metrics of the different sets of the Key Performance Indicators. The k-Mean clustering algorithm is implemented to show the clustered causes of network malfunctions.

## IV. CLASSIFICATION TECHNIQUES USING DECISION TREES

Retroactive data of 550 Base Station Controllers (BSC) were available. From the 550 BSC that were available, 440 cells were defined as training data and 110 were defined as test data. The following conditions and limits were set up for the optimisation system output:

@Attributes: 9
  @ Cell Performance
  @ Traffic Channel Call drop rate (numeric)
  @ Handover Success Seizure (numeric)
  @ SDCCH Drops (numeric)
  @ Radio Access Barrier (numeric)
  @ Handover Attempts (numeric)
  @ Handovers Failures Rate (numeric)
  @ Handover Success Rate (numeric)
  @ TCHDropSuddenLostCon
  @ Diagnosis (string)

The rules for the classification of causes generated from the decision tree are given bellow:

```
Rule-1: IF HandoverSussessSRate <= 71.23 and
         TCHCallDropRate <= 7.42 Handover failures <22.17
         THEN record is pre-classified as "Class A"
Rule-2: ELSE IF Handover failures >22.1 and RAB>3
         THEN symptom is pre-classified as "Class B"
         ELSE IF RAB< 3 THEN KPI is classified "Class A"
Rule-3: IF HandoverSussessSRate>57.99 and TCHCDR>1.18
         THEN record is classified as "Optimised"
         ELSE TCHCDR<=1.18 THEN record is classified as
         "Class C"
Rule-4: IF HandoverSussessSRate<= 57.99 and
         HandFailures>3.69 THEN record is "Class C"
Rule -5: IF HandFailures <3.69 and HandoverSussRate>25.15
         record is pre- classified as "Class A"
         ELSE HandoverSussessSRate<=25.15
         THEN record is classified as "Class C"
Rule -6: IF RAB <=6 and TCHCallDropRate <= 7.65
         THEN record is "Class C"
         ELSE TCHCallDropRate >7.65
         THEN  record is "Class A"
Rule-7: IF RAB >6 and TCHDSDLC>26
         THEN record is "Class B"
         ELSE TCHDSDLC<=26 and HandFailures <=2.87
         THEN record is "Class B"
Rule -8: IF HandFailures >2.87 and TCHCallDropR <= 3.18
         THEN record is "Class C"
         ELSE TCHCallDropR> 3.18
         THEN record is "Class C"
         END IF
```

Based on the results using the training data, an accuracy rate of 98.86% of instances were correctly classified. Only 6 instances were incorrectly classified. The mean absolute error was only 0.0087%. Table 1 shows the final statistics of the decision tree and the weighted averages for the decision tree. The results show that the weighted average precision of the algorithm was 98.64%, and that none of the classes had precision rates of less than 97%. Table 2 shows the confusion matrix, which shows that 93 causes of class A were correctly predicted, 216 out of 220 causes of class B were also predicted correctly and 64 out 65 causes are classified as class C. Only 61 causes of class D were predicted correctly. The optimisation decision tree derived from the results of the data analysis is shown in Fig. 3.

## V. IMPLEMENTATION OF DECISION TREE ALGORITHM

A decision tree has the shape of an inverted tree with a single point at the top, and branches that move down the tree. A decision tree is used to predict the value of a class, which is a target attribute, or labels that are based on input attributes of datasets. Each of the interior nodes of the decision tree is matched to an input attribute, known as a symptom. The number of potential values of each input attribute is the same as the number of edges of the interior node. The disjoint range label is assigned to outgoing edges of numerical attributes.

The results of this study show, as has been shown in other studies in which Weka was used to conduct experiments on network optimisation, the value and high level of accuracy rated to determining network optimisation using the machine learning language [11][16][17]. The J48 decision tree has shown where causes of network congestion can occur, as well as the symptoms that can create the lack of network optimisation so that engineers and operators can work to regain optimisation for end users. In addition, the results of this study have shown that the use of Weka provides a higher level of accuracy in determining network optimisation than methods used in other studies, including smooth Bayesian networks and drive testing and overhead mapping [7][15].

The advantage of using a decision tree is that the data contained within the tree is easy to read and interpret, and is meaningful to the user. Interpreting data in a decision tree is easier than in other approaches. However, the disadvantage of using a decision tree is that the causes of the optimisation diagnosis are not shown. Instead, only the values are shown with no causal information [23].

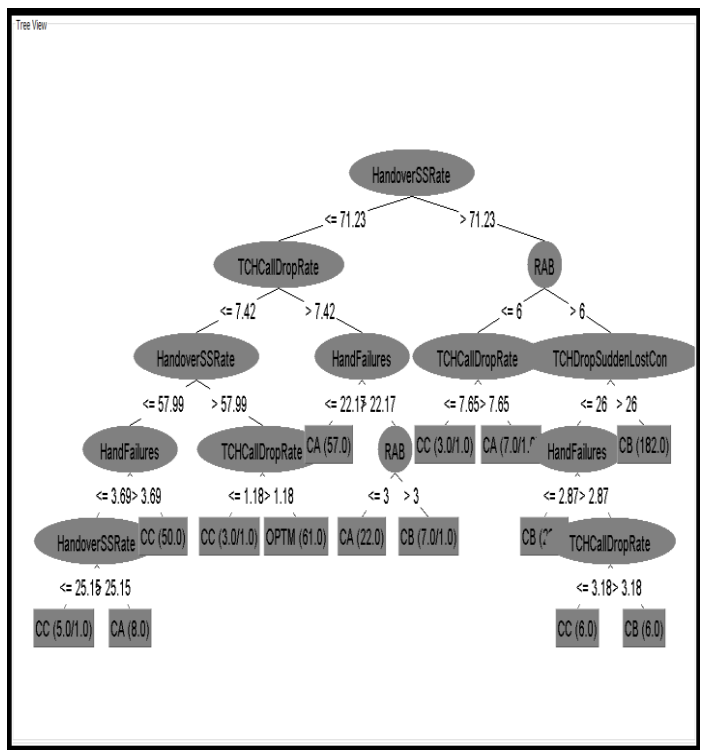

Figure 3. Optimisation decision tree

Even with the strong evidence from the results of this study for the use of the C4.5 classification algorithm to predict causes of problems in networks that result in a lack of optimisation, more work is needed. Further increases in accuracy might be achieved through the investigation of the solutions that influence performance of the network. The study of the performance metrics that are included in the troubleshooting management might also result in determining the parameters of the KPI faults that lead to a greater optimisation of a specific network.

The significance of this study is enhanced because of the use of clustering techniques in order to investigate optimisation of the mobile network. The use of the clustering techniques allows for the use of the multi-objective approach that has been missing in much of the previous research on this topic. The result will be a study that will improve upon and enhance the techniques that have been used in previous studies, while also providing new information and techniques that mobile network operators can use in actual practice.

VI. PROPOSED OPTIMISATION ALGORITHM

The clustering optimisation algorithm was based on the results of the decision tree classifier in which the limits of the different KPIs were identified and the values were used to set up the cause group. The proposed clustering algorithm will be asked to identify the areas of optimisation and to insert the individual values for the KPIs so that the results can be viewed in greater detail.

**Clustering Algorithm:**

Metrics Estimation for Multiple Technical Areas

**Input data:**

Training set of KPIs = {($x_i$, $y_i$, )} n i=1, Learning C45 algorithm that define an optimisation scheme based on the efficient KPI metrics according to the decision tree classification.

L to train the J48 classifier h : X → Y ,

**Step 1**: Estimate the different metrics that contribute to an optimisation scheme according to the decision tree classification.

**Step 2**: Define the group of causes Ca, Cb, Cc, Cd, present the KPI values of each network performance areas:

Ai ($T_N$, $D_N$, $H_N$, $CS_N$…$H_N$).

- Traffic= {$T_1$, · · · , $T_N$}.,
- Dropped Calls = {$D_1$, · · · , $D_N$},
- Handover Success = {$H_1$, · · · ,$H_N$},
- Call Successful Rate = {$C_1$, · · · , $C_N$}.

26

**Step 3**: Estimate Group of Causes (GCi) using the rules of the classification tree J48 algorithm,

**Step 4**: Identify the population of cells that used to determine the optimal set of group causes

**Step 5:** For each KPI value = 1, 2, 3,... ,i we carry out a clustering area and compute Attributes ($C_i$ = 0, 1, 2...i);

**Output:** Calculate the optimised cells for each cluster partition that shows the clustered network performance areas.

The clustering algorithm identifies the areas of the optimisation issues based on the value of the symptoms and causes of network malfunctions

## VII. ANALYSIS AND EXPERIMENTS

In order to evaluate the accuracy of the prediction of the k-means clustering with a decision tree as compared to the use of classifiers without a decision tree, a series of experiments were performed on data that were obtained from the network. The figure shows the characteristics of each KPI by analyzing the application of network optimisation with eight cluster centers [23].

The model can be run with a different number of attributes based on the characteristics of the mobile ad hoc network and the multidimensional queries that network operators request. In addition, the number of symptoms and the amount of cells can also be changed based on the technical areas of concern of the network operators. The number of symptoms and the amount of cells can also be changed based on the geographical areas of the network that are being examined, and the level of analysis that is desired by the network operators [2].

The process that was used involved introducing a small number of attributes. The reason for this is that with a smaller number of attributes, the results will be more precise and indicative for the technical area being examined. However, introducing more variables could provide the opportunity to examine the characteristics of 9 clusters that would represent the different technical areas of concern. If more variables were introduced into the model, the analysis would detail every area and provide a more general perspective of a possible optimisation solution. In such a situation, the combination of KPIs would not always be predictable.

TABLE I.  EXPERIMENT RESUTLS

| Results of K-Means Algorithm | | | | |
|---|---|---|---|---|
| **Attribute (KPI)** | *Cluster 0* | *Cluster 1* | *Cluster 2* | *Cluster 3* |
| **TCH Failures** | | | | |
| Mean | 2.32 | 13.51 | 22.83 | 7.55 |
| Std.Dev | 0.15 | 9.22 | 0.38 | 0.04 |
| **TCH Attempts** | | | | |
| Mean | 14.58 | 36.20 | 79.23 | 18.93 |
| Std.Dev | 0.27 | 27.03 | 0.38 | 0.03 |
| **RAB** | | | | |
| Mean | 0 | 0.032 | 123.71 | 101.30 |
| Std.Dev | 54.26 | 0.251 | 42.93 | 7.32 |
| **Handover Failures** | | | | |
| Mean | 2.82 | 11.30 | 31.12 | 10.56 |
| Std.Dev | 0.26 | 8.34 | 0.42 | 0.07 |
| **TCH Dropped Suddenly Lost Connection** | | | | |
| Mean | 0 | 0 | 130.45 | 97.13 |
| Std.Dev | 53.43 | 53.43 | 29.69 | 6.89 |
| **TCH Congestion Rate** | | | | |
| Mean | 2.82 | 11.30 | 31.12 | 10.56 |
| Std.Dev | 0.267 | 8.34 | 0.42 | 0.07 |
| **Handover Success Rate** | | | | |
| Mean | 61.11 | 48.73 | 73.23 | 51.34 |
| Std.Dev | 0.196 | 21.03 | 0.38 | 0.02 |

For example, Table I shows that in cluster 2, the results indicate unusual channel drops in which connection was suddenly lost. This occurred even though handover metrics and the traffic channel failures had reasonable and acceptable values. However, the cluster results show that the attempts might have been higher than normal, which could have affected the traffic of calls and data at that specific time. In that case, the next step would be to run the model including the traffic rate indicators to find out if the traffic equipment needs to be improved in that particular geographic area. If the results are also predictable, then other variables, such as peak times, would need to be checked that might affect the optimization plan.

The causes of the problem could also be clustered to show additional evidence as part of the optimisation results. It should be noted that one limitation of the model might be that causes for some of the groups may not be shown within some of the clusters. In this situation, the categories and clusters would have to be extended and further analysed.

___

TABLE II.     EXPERIMENT RESULTS

| Results of K-Means Algorithm | | | | | |
|---|---|---|---|---|---|
| Attribute (KPI) | Cluster 4 | Cluster 5 | Cluster 6 | Cluster 7 | Cluster 8 |
| TCH Failures | | | | | |
| Mean | 3.58 | 20.93 | 56.97 | 2.61 | 7.21 |
| Std.Dev | 0.44 | 7.80 | 16.99 | 0.43 | 0.09 |
| TCH Attempts | | | | | |
| Mean | 57.71 | 72.94 | 7639 | 75.95 | 18.60 |
| Std.Dev | 0.66 | 45.56 | 6169 | 9.30 | 0.09 |
| RAB | | | | | |
| Mean | 43.257 | 17.88 | 0.6 | 76.05 | 36.73 |
| Std.Dev | 25.42 | 9.62 | 2.24 | 36.98 | 18.26 |
| Handover Failures | | | | | |
| Mean | 5.20 | 26.85 | 65.33 | 2.73 | 9.90 |
| Std.Dev | 0.31 | 8.92 | 14.12 | 0.48 | 0.18 |
| TCH Dropped Suddenly Lost Connection | | | | | |
| Mean | 96.22 | 35.79 | 5.13 | 31.66 | 37.64 |
| Std.Dev | 16.96 | 22.33 | 19.20 | 18.99 | 16.83 |
| TCH Congestion Rate | | | | | |
| Mean | 5.20 | 26.85 | 65.33 | 2.73 | 9.90 |
| Std.Dev | 0.31 | 8.92 | 14.12 | 0.48 | 0.18 |
| Handover Success Rate | | | | | |
| Mean | 116.7 | 65.91 | 53.0 | 96.74 | 51.09 |
| Std.Dev | 4.22 | 13.72 | 15.77 | 1.76 | 0.07 |

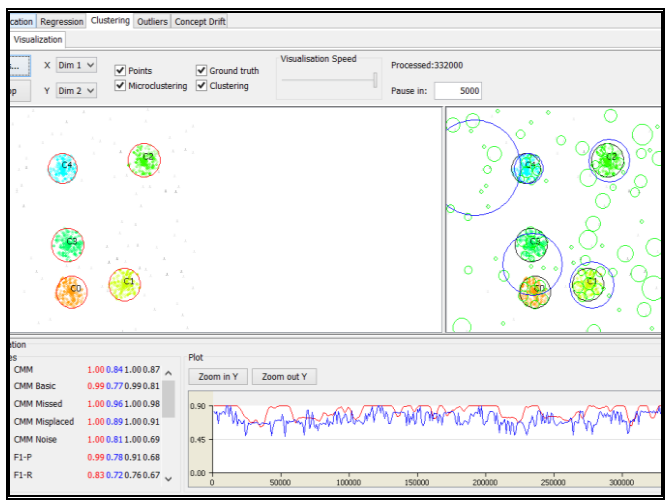

Figure 4. Evaluation of clustering metrics

28

___